\documentclass[aps,prc,twocolumn,letterpaper,floatfix]{revtex4}
\usepackage{graphicx,amsmath,amssymb}
\usepackage{url}

\begin{document}

\title{The Pilot-Wave Perspective on Quantum Scattering and Tunneling}
\author{Travis Norsen}
\affiliation{Smith College \\ Northampton, MA 01060 \\ tnorsen@smith.edu}

\date{December 14, 2012}

\begin{abstract}
The de Broglie - Bohm ``pilot-wave'' theory replaces the paradoxical
wave-particle duality of ordinary quantum theory with a more mundane
and literal kind of duality:  each individual photon or electron comprises
a quantum wave (evolving in accordance with the usual quantum mechanical wave
equation) \emph{and} a particle that, under the influence of the wave, 
traces out a definite trajectory.  The definite particle trajectory
allows the theory to account for the results of experiments without
the usual recourse to additional dynamical axioms about measurements.
Instead one need simply assume that particle detectors click when particles
arrive at them.  This alternative understanding of
quantum phenomena is illustrated here for two elementary textbook
examples of one-dimensional scattering and tunneling.  We introduce a novel
approach to reconciling standard textbook calculations (made using
unphysical plane-wave states) with the need
to treat such phenomena in terms of normalizable wave packets.  This
allows for a simple but illuminating analysis of the pilot-wave
theory's particle trajectories and an explicit demonstration of the
equivalence of the pilot-wave theory
predictions with those of ordinary quantum theory.  
\end{abstract}

\maketitle

\section{Introduction}

The pilot-wave version of quantum theory was originated in the
1920s by Louis de Broglie, re-discovered and developed in 1952 by
David Bohm, and championed in more recent decades especially by John
Stewart Bell. \cite{dbb} Usually described as a ``hidden variable'' theory, the
pilot-wave account of quantum phenomena supplements the usual
description of quantum systems -- in terms of wave functions -- with
definite particle positions that obey a deterministic evolution law.
This can be understood as the simplest possible account of
``wave-particle duality'':  individual particles (electrons,
photons, etc.) manage to behave sometimes like waves and sometimes
like particles because each one is literally both.  In, for example,
an interference experiment involving a single electron, the final
outcome will be a function of the position of the particle at the end
of the experiment.  (In short, detectors ``click'' when particles hit them.)
But the trajectory of the particle is not at all classical; it
is instead determined by the structure of the associated quantum wave
which guides or ``pilots'' the particle.

The main virtue of the theory, however, is not its deterministic
character, but rather the fact that it eliminates the need for
ordinary quantum theory's 
``unprofessionally vague and ambiguous'' measurement axioms.
\cite{bellqft}   Instead, in the
pilot-wave picture, measurements are just ordinary physical processes,
obeying the same fundamental dynamical laws as other processes.  In
particular, nothing like the infamous ``collapse postulate'' -- and the associated
Copenhagen notion that measurement outcomes are registered in some
separately-postulated classical world -- are needed.  The pointers, for
example, on laboratory measuring devices will end up pointing in
definite directions because they are made of particles -- and
particles, in the pilot-wave picture, always have definite positions.

In the ``minimalist'' presentation of the pilot-wave theory
(advocated especially by J.S. Bell), the guiding wave is simply the
usual quantum mechanical wave function $\Psi$ 
obeying the usual Schr\"odinger equation:
\begin{equation}
i \hbar \frac{\partial \Psi}{\partial t} = - \frac{\hbar^2}{2m}
\frac{\partial^2 \Psi}{\partial x^2} + V(x) \Psi.
\end{equation}
The particle position $X(t)$ evolves according to
\begin{equation}
\frac{dX}{dt} = \left. \frac{j}{\rho} \right|_{x=X(t)} 
\label{dBBvel}
\end{equation}
where 
\begin{equation}
j = \frac{\hbar}{2mi}(\Psi^* \frac{\partial}{\partial x} \Psi - \Psi
\frac{\partial}{\partial x} \Psi^*)
\label{jdef}
\end{equation}
(the usual quantum probability current) and  $\rho = |\Psi|^2$
(the usual quantum probability density) satisfy the continuity
equation
\begin{equation}
\frac{\partial \rho}{\partial t} + \frac{\partial j}{\partial x}  = 0.
\label{continuity}
\end{equation}
Here we consider the simplest possible case of a single spinless particle
moving in one dimension.  The generalizations for motion in 3D and
particles with spin are trivial:  $\partial/\partial x$ and $j$ become
vectors, and the wave function becomes a multi-component
spinor obeying the appropriate wave equation.  For a system of $N$
particles, labelled $i \in \left\{1, ..., N\right\}$,
the generalization is also straightforward, though it should be noted
that $\Psi$ -- and consequently also $\vec{j}_i$ and $\rho$ -- are in this case
functions on the system's configuration space.  The velocity of particle $i$ at
time $t$ is given by the ratio $\vec{j}_i / \rho$ evaluated at the complete
instantaneous configuration; thus in general the velocity of each
particle depends on the instantaneous positions of all other
particles.  The theory is thus explicitly non-local.  Bell, upon
noticing this surprising feature of the pilot-wave theory, was
famously led to
prove that such non-locality is a necessary feature of \emph{any} theory
sharing the empirical predictions of ordinary quantum
theory. \cite{belltheorem}

Although the fundamental dynamical laws in the
pilot-wave picture are deterministic, the theory exactly reproduces
the usual stochastic predictions of ordinary quantum mechanics.  This
arises from the assumption that, although the initial \emph{wave function}
can be controlled by the usual experimental state-preparation
techniques, the initial \emph{particle position} is \emph{random}.  In
particular, for an ensemble of identically-prepared quantum systems
having $t=0$ wave function $\Psi(x,0)$, it is assumed that the
initial particle positions $X(0)$ are distributed according to
\begin{equation}
P[X(0)\!=\!x] = |\Psi(x,0)|^2.
\label{QEH}
\end{equation}
This is called the ``quantum equilibrium hypothesis'' or QEH.
It is then a purely mathematical consequence of the already-postulated
dynamical laws for $\Psi$ and $X$ that the particle positions will
be $|\Psi|^2$ distributed for all times:
\begin{equation}
P[X(t) \!= \! x] = |\Psi(x,t)|^2,
\label{equiv}
\end{equation}
a property that has been dubbed the ``equivariance'' of the $|\Psi|^2$
probability distribution. \cite{dgz}
To see this, one need simply note that the probability distribution $P$
for an ensemble of particles moving in a velocity field $v(x,t)$
will evolve according to
\begin{equation}
\frac{\partial P}{\partial t} + \frac{\partial}{\partial x} ( v P ) = 0.
\end{equation}
Since $j$ and $\rho$ satisfy the continuity equation, 
it is then immediately clear that, for $v = j / \rho$, $P = \rho$ is a
solution.

Properly understood, the QEH can actually be derived from the basic
dynamical laws of the theory, much as the expectation that complex
systems should typically be found in thermal equilibrium can be
derived in classical statistical mechanics. \cite{dgz2}  For our purposes, though,
it will be sufficient to  simply take the QEH as
an additional assumption, from which it follows that the
pilot-wave theory will make the same predictions as ordinary quantum
theory for any experiment in which the outcome is registered by the
final position of the particle.  That the pilot-wave theory makes the
same predictions as ordinary QM for arbitrary measurements then
follows from the fact that, at the end of the day, such measurement
outcomes are also registered in the position of something:  think, for
example, of the flash on a screen somewhere behind a Stern-Gerlach
magnet, the position of a pointer on a laboratory measuring device, or
the distribution of ink droplets in \emph{Physical Review}. \cite{bellqft}

In the present paper, our goal is to illustrate all of these ideas by
showing in concrete detail how the pilot-wave theory deals with some
standard introductory textbook examples of one-dimensional quantum
scattering and tunneling.  This alternative perspective should be of
interest to students and teachers of this material, since it provides
an illuminating and compelling intuitive picture of these phenomena.
In addition, because the pilot-wave theory (for reasons we shall
discuss) forces us to remember that real particles should always
be described in terms of finite-length \emph{wave packets} -- rather
than unphysical plane-waves -- the methods to be developed provide a
novel perspective on ordinary textbook scattering theory as well.  In
particular, we describe a certain limit of the usual rigorous
approach to scattering \cite{scattering} in
which the specifically \emph{conceptual} advantages of working with
normalizable wave packets can be had without any \emph{computational}
overhead:  the relevant details about the packet shapes can be worked
out, in this limit, exclusively via intuitive reasoning involving the group velocity.

The remainder of the paper is organized as follows.  In the next
section, we review the standard textbook example of reflection and
transmission at a step potential, explaining in particular why the use
of plane-waves is particularly problematic in the pilot-wave picture
and then indicating how the usual plane-wave calculations can
be salvaged by thinking about wave packets with a certain special shape.
The
next section explores the pilot-wave particle trajectories in detail,
showing in particular how the reflection and transmission
probabilities can be computed from the properties of a certain
``critical trajectory'' \cite{uu} that divides the possible trajectories into
two classes:  those that transmit and those that reflect.  In the
following section we turn to an analysis of quantum tunneling through
a rectangular barrier from the pilot-wave perspective.  
A brief final section summarizes the results and situates the
pilot-wave theory in the context of other interpretations of the
quantum formalism.

\section{Schr\"odinger Wave Scattering at a Potential Step}

Let us consider the case of a particle of mass $m$ incident, from the
left, on the step potential
\begin{equation}
V(x) = \left\{ 
\begin{array}{lcc}
0 & \text{if} & x<0 \\
V_0 & \text{if} & x>0
\end{array}
\right. .
\end{equation}
where $V_0 > 0$.
The usual approach is to assume that we are dealing with a particle of
definite energy $E$ (which we assume here is $> V_0$) in which case we
can immediately write down an appropriate general solution to the
time-independent Schr\"odinger equation:
\begin{equation}
\psi(x) = \left\{
\begin{array}{lcc}
A e^{i k_0 x} + B e^{-i k_0 x} & \text{if} & x<0 \\
C e^{i \kappa_0 x} & \text{if} & x>0
\end{array}
\right.
\label{step-psi}
\end{equation}
where $k_0 = \sqrt{2mE/\hbar^2}$ and $\kappa_0 =
\sqrt{2m(E-V_0)/\hbar^2}$.  The $A$-term represents the incident wave
propagating to the right toward the barrier.  The $B$-term represents a
reflected wave propagating back out to the left.  The $C$-term
represents a transmitted wave.  And there is, by assumption, no
incoming (i.e., leftward-propagating) wave to the right of the
barrier.  

The transmission and reflection probabilities depend on the relative
amplitudes ($A$, $B$, and $C$) of the incident, reflected, and
transmitted waves.  By imposing continuity of $\psi(x)$ and its
derivative at $x=0$ (these conditions being required in order that the
above $\psi(x)$ satisfy the Schr\"odinger equation \emph{at} $x=0$)
one easily finds that
\begin{equation}
\frac{B}{A} = \frac{k_0 - \kappa_0}{k_0 + \kappa_0}
\label{BA}
\end{equation}
and
\begin{equation}
\frac{C}{A} = \frac{2k_0}{k_0 + \kappa_0}.
\label{CA}
\end{equation}
A typical textbook approach is then to calculate the probability
current in each region.  Plugging Equation \eqref{step-psi} into
Equation \eqref{jdef} gives
\begin{equation}
j = \left\{
\begin{array}{lcc}
\frac{\hbar k_0}{m} \left( |A|^2 - |B|^2\right) &\text{if} & x<0 \\
\frac{\hbar \kappa_0}{m} |C|^2 & \text{if} & x>0 
\end{array}
\right.
\end{equation}
which can be interpreted as follows.  For $x<0$ there is both an
incoming probability flux proportional to $k_0 |A|^2$ and an outgoing
flux proportional to $k_0 |B|^2$.  The reflection probability $R$ can be
defined as the ratio of these, so
\begin{equation}
P_R = \frac{j_{\text{reflected}}}{j_{\text{incident}}} =
\frac{k_0|B|^2}{k_0|A|^2} = \frac{|B|^2}{|A|^2} = \frac{(k_0-\kappa_0)^2}{(k_0+\kappa_0)^2}.
\label{P_R}
\end{equation}
Similarly, for $x>0$, there is an outgoing probability flux
proportional to $\kappa_0 |C|^2$.  The transmission probability $P_T$ can
be defined as the ratio of this flux to the incident flux, so
\begin{equation}
P_T = \frac{j_{\text{transmitted}}}{j_{\text{incident}}} = \frac{\kappa_0
  |C|^2}{k_0 |A|^2} = \frac{2 k_0 \kappa_0}{(k_0 + \kappa_0)^2}.
\label{P_T}
\end{equation}
This approach to calculating $P_R$ and $P_T$ is however somewhat unintuitive,
in so far as the wave function involved is a stationary state.  This
makes it far from obvious how to understand the mathematics as
describing an actual physical process, unfolding in time, in which a particle, initially
incident toward the barrier, either transmits or reflects.  The
situation is even more problematic, though, from the point of view of
the pilot-wave theory.  Here, the particle is supposed to have some
definite position at all times, with a velocity given by Equation
\eqref{dBBvel}.  But, with the wave function as given by
Equation \eqref{step-psi}, the probability current $j$ for $x<0$ is
positive (since $|A| > |B|$).  And of course $\rho$ is positive.  So it follows
immediately that, in the pilot-wave picture, the particle velocity is
necessarily positive:  if the particle is in the region $x<0$, it will be moving
to the right, toward the barrier.  It cannot possibly reflect!

It is easy to see, however, that this is an artifact of the use of unphysical
(unnormalizable) plane-wave states.  Many introductory textbooks mention in
passing the possibility of instead using finite wave packets to
analyze scattering. \cite{townsend}  Griffiths, for example, makes
the following characteristically eloquent remarks:
\begin{quote}
``This is all very tidy, but there is a sticky matter of principle
that we cannot altogether ignore:  These scattering wave functions are
not normalizable, so they don't actually represent possible particle
states.  But we know what the resolution to this problem is: We must
form normalizable linear combinations of the stationary states just as
we did for the free particle -- true physical particles are
represented by the resulting wave packets.  Though straightforward in
principle, this is a messy business in practice, and at this point it
is best to turn the problem over to a computer.'' \cite{griffiths}
\end{quote}
Griffiths goes on to characterize the ``peculiar''
fact ``that we were able to analyse a quintessentially time-dependent
problem...using \emph{stationary} states'' as a ``mathematical
miracle''. 
Some texts go a little further into this ``messy business'' and treat 
the problem of an incident
(typically, Gaussian) packet in some analytic detail. \cite{shankar}

\begin{figure}[t]
\begin{center}
\scalebox{.8}{
\includegraphics{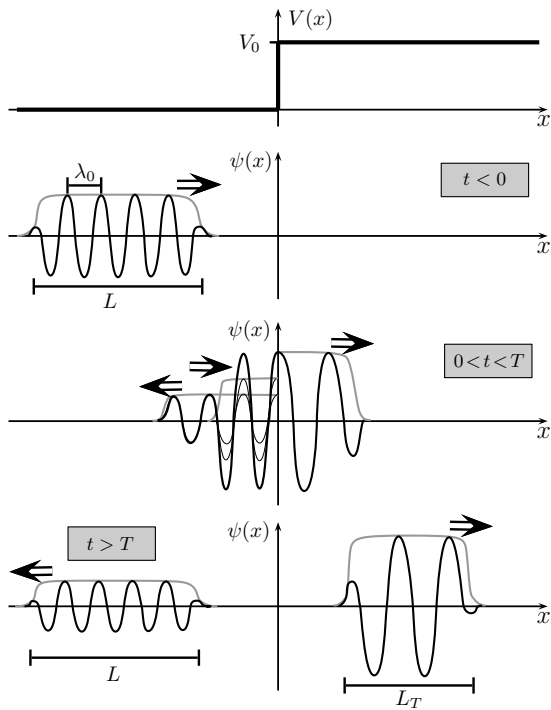}
}
\caption{Visualization of a ``plane-wave packet'' interacting with the
step potential shown in the top frame.  The ``$t<0$'' frame shows a
plane-wave packet of length $L$ and (central) wavelength $\lambda_0$
incident from the left.  (Note that, strictly speaking, a plane-wave
packet by definition has $L \gg \lambda_0$; the two length scales
are inappropriately similar in the figure so that several other features
will be more readily visible.)  At $t=0$ the leading edge of the
incident packet arrives at the origin and leading edges for the
reflected and transmitted packets are produced.  At $t=T$ the
trailing edge of the incident packet arrives at the origin and
trailing edges for the reflected and transmitted packets are
produced.  For $0 < t < T$ the incident and reflected packets
overlap in some (initially small, then bigger, then small again)
region to the left of the origin.  This is depicted in the
``$0<t<T$'' frame.  Note that the amplitudes of the reflected and
transmitted waves are determined by the usual boundary-matching
conditions imposed at $x=0$.  Finally, for $t>T$ the reflected and
transmitted packets propagate away from the origin.  Note that while
the wavelength and packet length of the reflected wave matches those
of the incident wave, the wavelength and packet length of the transmitted wave are
respectively \emph{greater} and \emph{smaller} than those of the
incident wave, owing to the different value of the potential energy to
the right of the origin. 
\label{fig1}
}
\end{center}
\end{figure}

And the need to examine scattering in terms of (finite, normalizable) 
wave packets has long been recognized in the pilot-wave literature,
which has included for example numerical studies of trajectories for
Gaussian packets incident on various barriers.  \cite{dh,uu,holland}
The use of Gaussian
packets, however, tends to obscure the relationship to the standard
textbook plane-wave calculation.  There is no way to express the
probabilities $P_R$
and $P_T$, for a narrow Gaussian packet, in anything like the
simple form of Equations \eqref{P_R} and \eqref{P_T}.  
And, in the
pilot-wave picture, the complicated structure of the wave function
during the scattering of the packet gives rise to equally complicated 
particle trajectories.  So although one of course knows based on
Equation \eqref{equiv} that the ensemble of possible particle trajectories will
``follow'' $\rho = |\psi|^2$, it is impossible to independently
\emph{verify} this without turning the problem over to a computer.

In the next section, we will develop a simple way to verify that,
indeed, just the right fraction of the possible particle trajectories
end up in the reflected and transmitted packets.  To lay the ground
for this, let us turn now to setting up a simple approach
to reconciling the plane-wave and wave packet approaches.
This should
be of pedagogical interest even to those with no particular interest
in the pilot-wave theory.  To be clear, what follows is in no sense
intended as a replacement for ordinary scattering
theory. \cite{scattering}  The point
is merely to show how, by considering incident packets with a
particular special shape, the reflection and transmission
probabilities can be read off in a trivial way from the packet
amplitudes and widths. 

Consider then an incident wave packet
\begin{equation}
\psi(x) = \phi(x) e^{i k_0 x}
\end{equation}
with a reasonably sharply-defined wave number $k_0$, but with a
special, non-Gaussian, envelope profile $\phi(x)$.  In particular, we
imagine $\phi(x)$ to be nearly \emph{constant} over a spatial region
of length $L$, and zero outside this region.  Then, as long as the
(central) wavelength $\lambda_0 = 2 \pi / k_0$ is very small compared to
$L$ (actually, it should in addition be small compared to the length
scale over which $\phi$ transitions to zero at the edges of the
packet) the envelope function $\phi$ will maintain its shape and
simply drift at the appropriate group velocity.  Let us call this type
of packet a ``plane-wave packet''; its conceptual and analytical merit
lies in the fact that, where it doesn't vanish, it is
well-approximated by a plane wave.  \cite{planewave}

In terms of such plane-wave packets, the scattering process can be
understood as shown in Figure \ref{fig1}.  Let us choose $t=0$ to be
the time when the leading edge of the incident packet arrives at
$x=0$.  The incident packet has length $L$
and moves at the group velocity $v_g^< = \hbar k_0 / m$.  Thus, the
packet's trailing edge arrives at the origin at $t = T = L/v_g^< = Lm/\hbar
k_0$.  The whole scattering process then naturally breaks up into the
following three time periods:
\begin{enumerate}
\item For $t<0$ the incident packet is propagating toward the barrier
  at $x=0$.
\item For $0 < t < T$ the wave function in some
  (initially small, then bigger, then small again) region around $x=0$
  is well-approximated by the plane-wave expressions of Equation
  \eqref{step-psi}.  
\item For $t > T$ the incident packet has completely
  disappeared and there are reflected and transmitted packets
  propagating away from the barrier on either side.
\end{enumerate}
It is now possible to understand the usual reflection and transmission
probabilities in a remarkably simple way.  To begin with, the
incident packet should be properly normalized.  Since it goes as $A
e^{i k_0 x}$ over a region of length $L$, it is clear that 
\begin{equation}
|A| = \frac{1}{\sqrt{L}}.
\end{equation}
The total probability associated with the reflected packet can be
found by multiplying its probability density $\rho_R = |B|^2$ by its
length.  Since the leading and trailing edges of the reflected packet
are produced respectively when the leading and trailing edges of
the incident packet arrive at the barrier -- and since the reflected
packet propagates in the same region as the incident packet so their
group velocities are the same -- the reflected packet has the same
length, $L$, as the incident packet.  Hence
\begin{equation}
P_R = |B|^2 L =\frac{|B|^2}{|A|^2} =
\frac{(k_0-\kappa_0)^2}{(k_0+\kappa_0)^2}
\end{equation}
where in the last step we have used Equation \eqref{BA} to relate the amplitude $B$ of
the reflected packet to the amplitude $A$ of the incident one. 
The result here is of course in agreement with Equation \eqref{P_R}.  

The transmission probability
can be calculated in a similar way.  But here it is crucial to recognize
that the group velocity for the $x>0$ region, $v_g^> = \hbar \kappa_0
/ m$, is \emph{smaller} than the group velocity in the $x<0$ region.
Thus, the position of the leading edge of the transmitted packet when
the trailing edge is created at $x=0$, i.e., the length of the
transmitted packet, is only
\begin{equation}
L_T = v_g^> \cdot T = v_g^> \cdot \frac{L}{v_g^<} = L \frac{\kappa_0}{k_0}.
\end{equation}
That is, the transmitted packet is \emph{shorter}, by a factor
$\kappa_0/k_0$, than the incident and reflected packets.  The total
probability carried by the transmitted packet is then easily seen to
be
\begin{equation}
P_T = L_T \, |C|^2 = L |C|^2 \frac{\kappa_0}{k_0} =
\frac{\kappa_0}{k_0} \frac{|C|^2}{|A|^2} =
\frac{4 k_0
  \kappa_0}{(k_0+\kappa_0)^2}
\end{equation}
again in agreement with the earlier result.  Note though, on this
analysis, how the perhaps-puzzling factor of $\kappa_0 / k_0$ in
Equation \eqref{P_T} admits an intuitively clear origin in the relative
lengths of the incident and transmitted packets.  

Even in the context of conventional, textbook quantum theory, the
``plane-wave packet'' approach has several pedagogical merits.  First,
it allows the scattering process to be understood and visualized as a 
genuine, time-dependent process.  Second, the reflection and
transmission probabilities can be calculated without recourse to the
somewhat cryptic and somewhat hand-waving device of taking ratios of
certain hand-picked terms from the probability currents on each side.
And finally, the explicit discussion of wave packets helps make clear
that the results of the calculation -- in particular the expressions
for $P_R$ and $P_T$ -- can be expected to be accurate only under the
conditions (e.g., $L \gg \lambda_0$) assumed in the derivation.  And
of course the over-arching point is that all of this is accomplished
while still using the mathematically simple plane-wave calculations:
there is no particularly ``messy business'' and no need ``to turn
the problem over to a computer.''

In the following section, we will see the particular utility of the
``plane-wave packet'' approach in the context of the alternative pilot-wave
picture.

\section{Particle Trajectories in the Pilot-Wave Theory}

In the pilot-wave theory, the particle velocity is determined by the
structure of the wave function in the vicinity of the particle,
according to Equation \eqref{dBBvel}.  By considering a plane-wave
packet as discussed in the previous section, we can see that there are
several possible regions in which the particle may find itself.  Let
us consider these in turn.

To begin with, initially, the particle will be at some (random)
location in the incident packet.  Since, by assumption, the packet
length $L$ is very large compared to the length scale associated with
the packet's leading and trailing edges, the particle is
overwhelmingly likely to be at a location where the wave function in
its immediate vicinity is given by
\begin{equation}
\psi_I(x) = A e^{ik_0x}.
\end{equation}
(Here and subsequently we omit for simplicity the time-dependent phase
of the wave function, which plays no role.)  It follows immediately
that the particle's velocity is
\begin{equation}
v_I = \frac{j_I}{\rho_I} = \frac{\frac{\hbar k_0}{m}
  |A|^2}{|A|^2} = \frac{\hbar k_0}{m}.
\end{equation}
Note that this is the same as the group velocity of the incident
packet.  Thus, the particle will approach the barrier with the incident
packet, indeed keeping its same position relative to the front and
rear of the packet as both things (the wave and the particle) move.

\begin{figure}[t]
\begin{center}
\scalebox{.9}{
\includegraphics{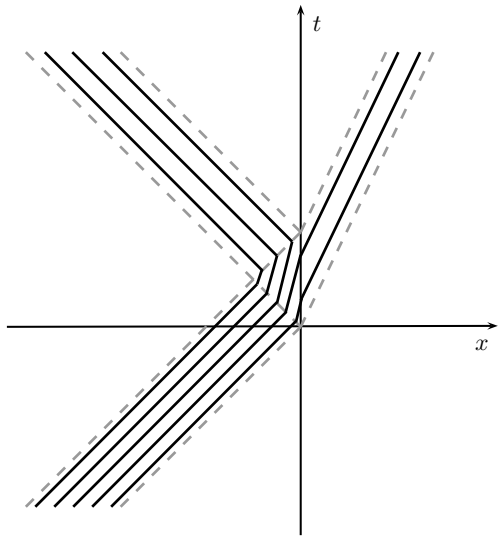}
}
\caption{
Space-time diagram showing a
representative sample of possible particle trajectories for the case
of a plane-wave packet incident from the left on a step potential.
The leading and trailing edges of the various packets are indicated by
the dashed grey lines.  Particle trajectories are shown in black.  In 
general, the particle simply moves at the group velocity along with
the packet that is guiding it.  In the (here, triangular) overlap
region, however, the particle moves more slowly.  This gives rise to a
bifurcation of the possible trajectories, between those which arrive at the
origin before being caught by the incident packet's trailing edge
(and thus end up moving away with the transmitted packet), and those
which are caught by the incident packet's trailing edge (and thus end
up moving away with the reflected packet).  
\label{fig2}
}
\end{center}
\end{figure}

At some point -- the exact time and place depending on its random
initial position within the incident packet -- the particle will
encounter the leading edge of the reflected packet.  It will then
begin to move through the ``overlap region'' where both the incident
and reflected waves are present:
\begin{equation}
\psi_O(x) = A e^{ik_0x} + B e^{-ik_0x}.
\end{equation}
Its velocity in this overlap region will be given by
\begin{equation}
v_O = \frac{j_O}{\rho_O} = \frac{ \frac{\hbar k_0}{m} \left(
    |A|^2 - |B|^2 \right)}{|A|^2 + |B|^2 + 2 |A|  |B| \cos(2 k_0 x-\phi)}
\label{vO}
\end{equation}
where $\phi$ is the complex phase of $B$ relative to $A$ -- zero in
the case at hand.
Here the right hand side is to be evaluated at each moment at the
instantaneous location of the particle.  This first-order differential
equation for $X(t)$ is easily solved -- more precisely, it is trivial to
write an exact expression for $t(X)$ -- but it is already clear from
the above expression that the particle's velocity will oscillate
around an average ``drift'' value given by
\begin{equation}
\bar{v}_O = \frac{\hbar k_0}{m} \frac{ |A|^2 - |B|^2}{|A|^2 + |B|^2}.
\label{voverlap}
\end{equation}
Since we are assuming that the packet length $L \gg \lambda_0 = 2 \pi
/ k_0$, the particle's velocity will (with overwhelming probability)
oscillate above and below this average value many many times while it
moves through the overlap region.  It is
thus an excellent approximation to simply ignore the oscillations and
treat the particle as moving through the overlap region with a
constant velocity $\bar{v}_O$.

There are two possible ways for the particle to escape from the
overlap region.  If it arrives at the origin, it will cross
over into the region where only the transmitted wave
\begin{equation}
\psi_T(x) = C e^{i \kappa_0 x}
\end{equation}
is present.  It will then continue to move to the right with a
velocity
\begin{equation}
v_T =\frac{j_T}{\rho_T}= \frac{\hbar \kappa_0}{m}
\end{equation}
matching the group velocity of the transmitted packet.  

The other possibility is that, while still in the overlap region, the
trailing edge of the incident packet catches and surpasses the
particle.  It will then subsequently be guided exclusively by the
reflected wave
\begin{equation}
\psi_R(x) = B e^{-i k_0 x}.
\end{equation}
Its velocity
\begin{equation}
v_R = \frac{j_R}{\rho_R} = - \frac{\hbar k_0}{m}
\end{equation}
will match that of the reflected wave packet, with which it will
propagate back out to the left.

It is helpful to visualize the family of possible particle
trajectories on a space-time diagram; see Figure \ref{fig2}.  Notice
that a particle which happens to begin near the leading edge of the
incident packet will definitely transmit, while particles beginning
nearer the trailing edge of the incident packet will definitely
reflect.  

Although the dynamics here is completely deterministic, the
theory makes statistical predictions because the initial position of a
particular particle within its guiding wave is uncontrollable and
unpredictable.  Recall the ``quantum equilibrium hypothesis'' (QEH)
according to which, for an ensemble of identically-prepared systems
with initial wave function $\Psi(x,0)$, the initial particle positions
will be random, with distribution given by Equation
\eqref{QEH}.  It then follows, from the ``equivariance'' property
described in the introduction, that $\rho = |\psi|^2$ will continue to
describe the particles' probability distribution for all $t$.  The
pilot-wave theory thus reproduces the exact statistical predictions of
ordinary QM, but without any further axioms about measurement:
whereas in ordinary QM, for example, the transmission probability $P_T$
(equal to the integral of $\rho$ across the transmitted packet)
represents only the probability that the particle will appear there if
a measurement is made, in the pilot-wave theory $P_T$ instead represents
the probability that the particle \emph{really is there} in the
transmitted packet, ready to trigger a ``click'' in a detector should
such a device happen to be present.

\begin{figure}[t]
\begin{center}
\scalebox{1.0}{
\includegraphics{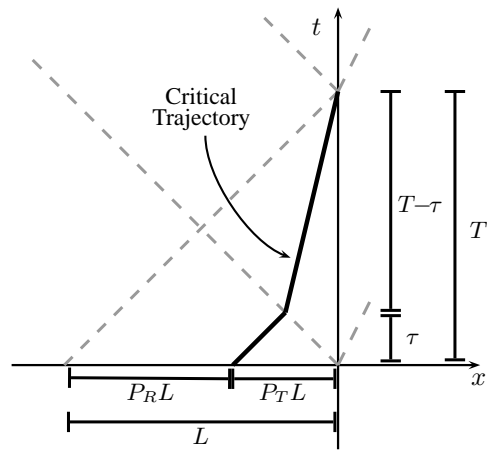}
}
\caption{
The critical trajectory, which arrives at the apex of the triangular overlap
region on this space-time diagram, 
divides trajectories that transmit from those that reflect.
The possible trajectories are distributed with uniform
probability density throughout the incident packet, so the fraction of
the total length $L$ of the packet that is in front of the critical
trajectory represents the transmission probability, $P_T$.
Equivalently, the critical trajectory is a distance $P_T\cdot L$ behind
the incident packet's leading edge.  From $t=0$, it covers exactly
half this distance before encountering the leading edge of the
reflected packet.  This occurs after a time $\tau = (P_T\cdot L /
2)/(\hbar k_0 / m)$.  In traversing the overlap region, the critical
trajectory then moves through the remaining distance $P_T \cdot L / 2$.
The time it takes is $T - \tau$, where $T = L/(\hbar k / m)$ is the
time needed for the trailing edge of the incident packet to arrive at
the origin.  Equation \eqref{vOtraj} in the main text follows
immediately by dividing this distance by this time.  
\label{fig3}
}
\end{center}
\end{figure}

As a concrete illustration of the equivariance property that guarantees
the equivalence between the pilot-wave theory's statistical
predictions and those of ordinary quantum theory, let us derive the
reflection and transmission probabilities directly from the particle
trajectories and show that we indeed arrive again at the same
expressions we found before by considering only the quantum wave.  The
key here is to examine in particular the ``critical trajectory'' which
divides those trajectories which result in transmission, from those
which result in reflection.  This critical trajectory, by definition,
arrives just at the apex of the triangular overlap region of Figure
\ref{fig2}:  particles on the leading-edge side of the critical
trajectory will necessarily transmit, while particles on the
trailing-edge side of the critical trajectory will necessarily
reflect.

A zoomed-in image of the overlap region from Figure \ref{fig2} is
shown in Figure \ref{fig3}.   As explained in the caption, the
critical trajectory moves through the overlap region across a distance 
$P_T\cdot L / 2$ where $P_T$ is the transmission probability.  It does
this in a time $T - \tau$ where $T = L m / \hbar k_0 $ and $\tau = (P_T
\cdot L / 2)/(\hbar k_0 /m)$.  It follows that the (average) velocity
through the overlap region is
\begin{equation}
\bar{v}_O = \frac{ P_T L / 2}{\frac{Lm}{\hbar k_0} - \frac{P_T L m}{2 \hbar
    k_0}} = \frac{\hbar k_0}{m} \frac{P_T}{2-P_T}.
\label{vOtraj}
\end{equation}
Equating this with the expression for the velocity in the overlap
region worked out in Equation \eqref{voverlap} gives
\begin{equation}
\frac{\hbar k_0}{m} \frac{P_T}{2-P_T} = \frac{\hbar k_0}{m} \frac{|A|^2 -
  |B|^2}{|A|^2 + |B|^2}
\end{equation}
which can be solved for $P_T$ to give
\begin{equation}
P_T = \frac{|A|^2 - |B|^2}{|A|^2}.
\end{equation}
Using Equation \eqref{BA} to put this in terms of the wave numbers
$k_0$ and $\kappa_0$ of course gives back precisely Equation \eqref{P_T} for the
transmission probability.  And since $P_R = 1-P_T$, 
Equation \eqref{P_R} is also implied again by the properties of the
critical trajectory.  

It is of course no surprise that we arrive at the same expressions for
the transmission and reflection probabilities by considering the
pilot-wave expression for the particle velocity in, especially, the
crucial overlap region.  But it is a clarifying confirmation of the
sense in which the wave and particle evolutions are \emph{consistent}
as expressed in the equivariance property.

\section{Tunneling through a rectangular barrier}

To illustrate the more general applicability of the methods developed
in the previous sections, let us analyze another standard textbook
example -- the tunneling of a particle through a classically forbidden
region -- from the pilot-wave perspective.  Let the potential be given
by
\begin{equation}
V(x) = \left\{
\begin{array}{lc}
V_0 & \text{if }  0<x<a \\
0 & \text{otherwise}
\end{array} 
\right.
\end{equation}
and let the particle, with a reasonably sharply defined energy $E <
V_0$, be incident from the left.  As before, we take the initial wave
function $\Psi(x,0)$ to be a plane-wave packet with (central)
wavelength $\lambda_0 = 2 \pi / k_0$ (with $k_0 = \sqrt{2mE/\hbar^2}$)
and length $L \gg \lambda_0$.  In addition, we assume here that the
packet length $L$ is much greater than the width $a$ of the potential
energy barrier.  Then -- letting now $\kappa_0 = \sqrt{2m (V_0 - E) /
  \hbar^2}$ -- the wave function in the vicinity of the barrier
will be given by
\begin{equation}
\psi(x) = \left\{
\begin{array}{lcc}
A e^{ik_0x} + B e^{-i k_0 x} & \text{if} & x < 0 \\ 
C e^{-\kappa_o x} + D e^{\kappa_0 x} & \text{if} & 0 < x < a \\
F e^{i k_0 x} & \text{if} & x > a 
\end{array}
\right.
\end{equation}
for the overwhelming majority of the time when $\psi(x)$ near the
barrier is nonzero.  (In particular, $\psi(x)$ will differ
substantially from the above expressions just when the leading edge of
the incident packet first arrives at the barrier, and again when the
trailing edge arrives there.  But this will have negligible effect on
our analysis since the probability for the particle to be too near the
leading or trailing edges will be, for very large $L$, very small.)  

Imposing the usual continuity conditions on $\psi(x)$ and its first
derivative at $x=0$ and $x=a$ gives a set of four algebraic conditions
on the amplitudes $A$, $B$, $C$, $D$, and $F$.  Eliminating $C$ and
$D$ allows the amplitudes of the reflected ($B$) and transmitted ($F$)
packets to be written in terms of the amplitude $A$ of the incident
packet:
\begin{equation}
\frac{B}{A} = \frac{ - (\kappa_0^2 + k_0^2) \sinh(\kappa_0 a)
}{(\kappa_0^2 - k_0^2) \sinh(\kappa_0 a) - 2 i k_0 \kappa_0
  \cosh(\kappa_0 a)}
\label{tunnelBA}
\end{equation}
and
\begin{equation}
\frac{F}{A} = \frac{-2i k_0 \kappa_0}{(\kappa_0^2 - k_0^2)
  \sinh(\kappa_0 a) - 2 i k_0 \kappa_0 \cosh(\kappa_0 a)} e^{-i k_0 a}.
\end{equation}
Since the packet that develops on the downstream side of the barrier
moves with the same group velocity as the incident packet, the
transmitted packet length matches the incident packet length.  The
total probability associated with the transmitted packet -- the
``tunneling probability'' -- is thus
\begin{equation}
P_T = \frac{|F|^2}{|A|^2} = \frac{4 k_0^2
  \kappa_0^2}{(\kappa_0^2+k_0^2)^2 \sinh^2(\kappa_0 a) + 4 k_0^2 \kappa_0^2}
\end{equation}
with the corresponding reflection probability being
\begin{equation}
P_R = \frac{|B|^2}{|A|^2} = \frac{ (\kappa_0^2 + k_0^2)^2
  \sinh^2(\kappa_0 a) }{(\kappa_0^2 + k_0^2)^2
  \sinh^2(\kappa_0 a)+ 4 k_0^2 \kappa_0^2}.
\end{equation}
As before, these results can be understood in terms of the particle
trajectories as well.  In general the trajectories are very similar to
those from the earlier example.  While the incident and reflected
packets are both present to the left of the barrier, an overlap region 
is set up in which the motion of the incoming particle is slowed.
The particle velocity in this region is again described by Equation
\eqref{vO}, although now there is a nontrivial complex phase between
the amplitudes $B$ and $A$:
\begin{equation}
\phi = \tan^{-1}\left( \frac{ 2 k_0 \kappa_0 \cosh(\kappa_0 a)
  }{(\kappa_0^2-k_0^2) \sinh(\kappa_0 a) } \right).
\end{equation}
The average drift velocity through the overlap region, however,
remains as in Equation \eqref{voverlap}, so the analysis surrounding
Figure \ref{fig3} still applies and we have again that the
transmission (or here, tunneling) probability as determined by the
critical trajectory is
\begin{equation}
P_T = \frac{|A|^2 - |B|^2}{|A|^2}
\end{equation}
in agreement with the result arrived at by considering just the
waves.  This again confirms that the distribution of possible particle
trajectories evolves in concert with the wave intensity $\rho$ such
that Equation \eqref{equiv} remains true at all times.  

\begin{figure}[t]
\begin{center}
\scalebox{.9}{
\includegraphics{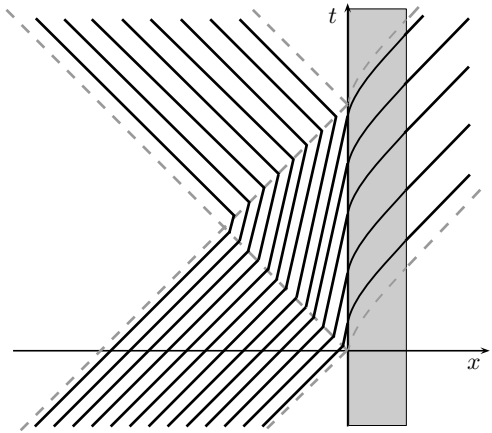}
}
\caption{
Space-time diagram showing a 
representative sample of possible particle trajectories for the case
of a plane-wave packet incident from the left on a rectangular
potential barrier.  Particles beginning near the leading edge of the
incident packet will tunnel through the barrier and emerge on the far
side.  The unusual accelerating character of the trajectories in the
(here, grey shaded) classically-forbidden region -- indeed the mere
presence of trajectories here -- reflects the highly non-classical
nature of the law of motion for the particle.  (Note that the analysis
in the main text assumes that the incident packet length $L$ is very
large compared to the barrier width $a$.  This separation of length
scales is not accurately depicted in the Figure, in order that the
qualitative nature of the trajectories in all relevant regions can be
visualized simultaneously.)
\label{fig4}
}
\end{center}
\end{figure}

The nature of the pilot-wave theory particle 
trajectories in the classically-forbidden region 
(CFR) is of some interest.  The wave function in the
CFR goes as
\begin{equation}
\psi_{\text{CFR}}(x) = C e^{-\kappa_0 x} + D e^{\kappa_0 x}
\end{equation}
where the four algebraic conditions mentioned
just prior to Equation \eqref{tunnelBA} imply that 
\begin{equation}
\frac{D}{C} = \frac{(\kappa_0 + i k_0)^2}{\kappa_0^2 + k_0^2} e^{-2
  \kappa_0 a} = e^{i \theta} e^{-2 \kappa_0 a}
\end{equation}
where the relative phase $\theta$ is given by
\begin{equation}
\theta = 2 \tan^{-1}(k_0/\kappa_0).
\end{equation}
The fact that the relative complex phase of $D$ and $C$ is not zero is
crucial:  if this were zero, the probability current $j$ and hence the
particle velocity would vanish, and it would be impossible for the
particles to tunnel across the barrier.  Instead, though, we have that
\begin{equation}
j_{\text{CFR}}(x) = \frac{2 \hbar \kappa_0}{m} |C|^2 \sin(\theta)
e^{-2 \kappa_0 a}
\end{equation}
and
\begin{equation}
\frac{\rho_{\text{CFR}}(x)}{|C|^2} =  e^{-2 \kappa_0 x} + e^{-4 \kappa_0 a} e^{2 \kappa_0 x} + 2 e^{-2 \kappa_0 a} \cos(\theta) 
\end{equation}
so that the particle velocity is given by
\begin{equation}
v_{\text{CFR}}(x) = \frac{\hbar \kappa_0}{m} \frac{ \sin(\theta)
}{\cos(\theta) + \cosh\left[ 2 \kappa_0 (a-x)\right]}.
\end{equation}
Thus, particles \emph{speed up} as they cross (from $x=0$ to $x=a$)
over the CFR.  

Figure \ref{fig4} displays the behavior of a representative 
sample of particle trajectories.  The overall pattern is similar to
the case of scattering from the step potential:  particles that begin
near the trailing edge of the incident packet will be swept up by the
reflected packet before reaching $x=0$, while those which begin nearer
the leading edge of the incident packet will reach the barrier, tunnel
across it, and emerge with the transmitted packet.

\section{Discussion}

We have analyzed two standard textbook cases of one-dimensional
quantum mechanical scattering and tunneling from the point of view of the de
Broglie - Bohm pilot-wave theory.  In particular, we have shown how
the standard textbook expressions for the reflection and
transmission/tunneling probabilities -- calculated using
infinitely-extended plane-wave states -- can instead be understood as arising
from a certain type of idealized, non-Gaussian incident wave packet.  We then
took advantage of this ``plane-wave packet'' approach to generate a
tractable, indeed quite simple, picture of how the particle
trajectories in the pilot-wave theory develop.  

It is hoped that the plane-wave packet approach might prove
clarifying for students learning standard textbook quantum mechanics.
It is also hoped that introducing the pilot-wave theory
through standard textbook examples will make it easier for teachers to
present the range of available interpretive options clearly and effectively to
students.  Recent work has shown that modern physics students have particular
difficulty with conceptual questions involving issues of
interpretation \cite{wuttiprom} -- a finding that is hardly surprising
given that physics teachers themselves have divergent views on
interpretive questions and their place in the curriculum.
\cite{baily}  These questions deserve to be discussed more explicitly
and more carefully, and it seems natural to do so in the context of
the kinds of example problems that students encounter in such courses
anyway.  

Despite its not being suggested as an option in the textbook or
lectures, several of the students interviewed in
Ref. \onlinecite{baily} seem to have independently developed a
pilot-wave type understanding of single-particle interference
phenomena.  Many eminent physicists have also found a pilot-wave
ontology to be the natural way to account for puzzling quantum
effects.  Here, for example, is J.S. Bell on 
single-particle interference experiments:
\begin{quote}
``While the founding fathers agonized over the question 
\begin{center}
`particle' \emph{or} `wave' 
\end{center}
de Broglie in 1925 proposed the obvious answer
\begin{center}
`particle' \emph{and} `wave'.
\end{center}
Is it not clear from the smallness of the scintillation on the screen
that we have to do with a particle?  And is it not clear, from the
diffraction and interference patterns, that the motion of the particle
is directed by a wave?  De Broglie showed in detail how the motion of
a particle, passing through just one of two holes in [the] screen,
could be influenced by waves propagating through both holes.  And so
influenced that the particle does not go where the waves cancel out,
but is attracted to where they cooperate.  This idea seems to me so
natural and simple, to resolve the wave-particle dilemma in such a
clear and ordinary way, that it is a great mystery to me that it was
so generally ignored.'' \cite{bell6}
\end{quote}
In an earlier paper, Bell asked:
\begin{quote}
``Why is the pilot
wave picture ignored in text books? Should it not be taught, not as
the only way, but as an antidote to the prevailing complacency? To 
show that vagueness, subjectivity, and indeterminism are not forced 
on us by experimental facts, but by deliberate theoretical 
choice?'' \cite{bellwhy}
\end{quote}
If current physicists answered these questions, the majority would probably
cite two factors, both of which involve some confusion and
mis-information.  First, there is the oft-repeated charge that the
pilot-wave theory involves an \emph{ad hoc} and cumbersome additional
field -- the so-called ``quantum potential'' -- to guide the
particle.  The theory has indeed been presented in such a form by Bohm
and others. \cite{uu,holland}  But as the examples in the body of the
present work should help make clear, this is an entirely unnecessary
addition to the ``minimalist'' pilot-wave theory, in which the field
guiding the particle is none other than the usual quantum mechanical
wave function obeying the usual Schr\"odinger equation.  

The second factor typically cited by critics of the pilot-wave theory
is its non-local character and the associated alleged incompatibility
with relativity. 
It is true, as discussed just after Equation \eqref{continuity} above,
that the pilot-wave theory is explicitly non-local.  What the critics
forget, however, is that ordinary quantum mechanics is also a
non-local theory:  already in its account of the simple one-particle 
scattering phenomena discussed here, orthodox quantum theory needs
additional postulates -- in particular the infamous and manifestly
non-local \emph{collapse postulate} -- to explain what is empirically
observed.  The truth is that, as we know from Bell, no local theory
can be empirically adequate.  \cite{belltheorem}  So rejecting candidate
interpretations on the basis of their non-local character is
hardly appropriate.  Nevertheless, it is interesting that the
conventional wisdom on this point is completely backwards:  
the pilot-wave theory is actually \emph{less} non-local than ordinary
quantum theory in the sense that it (unlike the orthodox theory) can
at least account for the results of one-particle
scattering/tunneling/interference experiments in a completely local
way.  

It is thus hoped not only that the examples presented here will
provide a simple concrete way for the alternative pilot-wave picture
to be introduced to students, but also that the
examples will help overturn some unfortunate and widely-held 
misconceptions about the theory.  And of course it should be noted
that the pilot-wave theory is just one of several alternatives to the
usual Copenhagen-inspired theory that appears in most textbooks.
There is, for example, also the many-worlds (``Everettian'') theory,
the spontaneous collapse (``GRW'') theory, the consistent (or
decoherent) histories approach, and many others.   As
someone who thinks that these questions -- about the \emph{physics}
behind the quantum formalism -- are meaningful, important,
fascinating, controversial, and too-often hidden under a shroud of
unspeakability, I would like to see all of these interpretations more
widely understood and discussed by physicists, both in and out of the
classroom.  (Some suggestions for introducing the issues and options
to students can be found in Ref. \onlinecite{suggestions}.)
At the end of the day, though, I cannot help but agree with Bell, who,
after reviewing ``Six possible worlds [i.e., interpretations] of
quantum mechanics'', concluded that ``the pilot wave picture
undoubtedly shows the best craftsmanship.''  \cite{bell6}  Hopefully
the examples discussed above will help others appreciate why.

\begin{acknowledgments}
Thanks to Shelly Goldstein, Doug Hemmick, and two anonymous referees
for helpful comments on earlier drafts of the paper. 
\end{acknowledgments}


\begin{thebibliography}{99}

\bibitem{dbb}  An English translation of Louis de Broglie's 1927
  pilot-wave theory can be found in G. Bacciagalluppi and
  A. Valentini, \emph{Quantum Theory at the Crossroads}, Cambridge,
  2009.  David Bohm's 1952 re-discovery of the theory is presented in
  ``A Suggested Interpretation of the Quantum Theory in terms of
  Hidden Variables, I and II'', \emph{Physical Review}, 85 (1952),
  pp. 166-193.  A more contemporary overview, with further references,
  can be found at \url{plato.stanford.edu/entries/qm-bohm}.

\bibitem{bellqft}  J.S. Bell, ``Beables for Quantum Field Theory''
  (1984) in \emph{Speakable and Unspeakable in Quantum Mechanics}, 2nd
  ed., Cambridge, 2004. 

\bibitem{scattering} See for example R.G. Newton, \emph{Scattering
    Theory of Waves and Particles}, 2nd Ed., Springer, Berlin /
  Heidelberg / New York, 1982; M. Reed and B. Simon, \emph{Methods of
    Modern Mathematical Physics III: Scattering Theory}, San Diego,
  Academic Press, 1979.

\bibitem{belltheorem}  J.S. Bell, ``On the Einstein-Podolsky-Rosen
  Paradox'', \emph{Physics} 1 (1964) 195-200.  Reprinted in Bell,
  2004, op cit.  For a contemporary systematic review of Bell's
  Theorem, see S. Goldstein, T. Norsen, D. Tausk, and N. Zanghi,
  ``Bell's Theorem'' at \url{scholarpedia.org/article/Bell\%27s_theorem}

\bibitem{dgz} D. D\"urr, S. Goldstein, and N. Zanghi, ``Quantum
  Equilibrium and the Origin of Absolute Uncertainty'', \emph{Journal
    of Statistical Physics} {\bf{67}}, pp. 843-907.

\bibitem{dgz2} \emph{Ibid.}  See also M.D. Towler, N.J. Russell, and
  A. Valentini, ``Time scales for dynamical relaxation to the Born
  rule'', \emph{Proceedings of the Royal Society A}, 468 (2012), pp. 990-1013.

\bibitem{townsend} J.S. Townsend, \emph{Quantum Physics:  A
    Fundamental Approach to Modern Physics}, University Science Books,
  Sausalito, California, 2010.

\bibitem{griffiths}  D.J. Griffiths, \emph{Introduction to Quantum
   Mechanics}, Prentice Hall, New Jersey, 1995, p. 58 .


\bibitem{shankar} R. Shankar, \emph{Principles of Quantum Mechanics},
  Second Edition, Springer, 1994.

\bibitem{dh} C. Dewdney and B.J. Hiley, ``A Quantum Potential
  Description of One-Dimensional Time-Dependent Scattering From Square
  Barriers and Square Wells'' \emph{Foundations of Physics}, Vol. 12,
  1982, pp. 27-48.  

\bibitem{uu} D. Bohm and B.J. Hiley, \emph{The Undivided Universe},
  Routledge, London and New York, 1993, pp. 73-8.

\bibitem{holland} P. Holland, \emph{The Quantum Theory of Motion},
  Cambridge University Press, 1993, pp. 198-203 (and references therein).


\bibitem{planewave}  For a rather different (alternative) approach to
  scattering that also uses the terminology ``plane-wave packet'' see
  Stuart C. Althorpe, ``General time-dependent formulation of quantum
  scattering theory'', \emph{Physical Review A}, 69, 042702 (2004).

\bibitem{wuttiprom} S. Wuttiprom, M. D. Sharma, I. D. Johnston, R. Chitaree, and C. Soankwan, ``Development and Use of a Conceptual Survey in Introductory Quantum Physics'', \emph{Int. J. Sci. Educ.} {\bf{31}}, 631-654 (2009).

\bibitem{baily} C. Baily and N. Finkelstein, ``Refined
  characterization of student perspectives on quantum physics'', 
\emph{Phys. Rev. ST Physics Ed. Research} {\bf{6}}, 020113, pp. 1-11 (2010)

\bibitem{bell6} J.S. Bell, ``Six Possible Worlds of Quantum
  Mechanics,'' 1986, reprinted in Bell, 2004, \emph{op cit.}  (Note:
  ellipsis in original.)  

\bibitem{bellwhy} J.S. Bell, ``On the impossible pilot wave,'' 1982,
  reprinted in Bell, 2004, \emph{op cit.}

\bibitem{suggestions}  Bell's paper, ``Six Possible Worlds of Quantum
  Mechanics'', \emph{op. cit.}, reviews the relevant phenomena of
  single-particle interference and then surveys six different extant
  interpretations.  It is an extremely accessible introduction to the
  nature (and existence) of the controversies, and I have often used
  it as the basis for a one-class-period discussion in a
  sophomore-level modern physics course.  At a slightly more technical
  level, Sheldon Goldstein's two-part \emph{Physics Today} article
  [March, 1998, pp. 42-46 and April, 1998, pp. 38-42] gives an
  extremely clear presentation of three attempts to formulate a
  ``Quantum Theory Without Observers'' and an illuminating explanation
  for why such a thing should be desirable in the first place.  For
  students who want to explore foundational issues (EPR, Bell, etc.)
  and their emerging applications (quantum cryptography, computation,
  etc.) in more depth, I would recommend GianCarlo Ghirardi's lucid
  book, \emph{Sneaking a Look at God's Cards}, Revised Edition, Gerald
  Malsbary, trans., Princeton University Press, 2005.  

\end{thebibliography}
\end{document}